\documentclass[12pt]{article}
\usepackage[english,german,french,polish]{babel}
\usepackage[T1]{fontenc}

\begin{document}
\selectlanguage{english}

\baselineskip 0.76cm
\topmargin -0.4in
\oddsidemargin -0.1in

\let\ni=\noindent

\renewcommand{\thefootnote}{\fnsymbol{footnote}}

\newcommand{\SM}{Standard Model }

\pagestyle {plain}

\setcounter{page}{1}



~~~~~~
\pagestyle{empty}

\begin{flushright}
IFT-- 06/6
\end{flushright}

\vspace{0.4cm}

{\large\centerline{\bf Mass formula for leptons and quarks }}

\vspace{0.2cm}

{\large\centerline{\bf as suggested by generalized Dirac's square root{\footnote{Work supported in part by the Polish Ministry of Education and Science, grant 1 PO3B 099 29 (2005-2007). }}}}

\vspace{0.5cm}

{\centerline {\sc Wojciech Kr\'{o}likowski}}

\vspace{0.23cm}

{\centerline {\it Institute of Theoretical Physics, Warsaw University}}

{\centerline {\it Ho\.{z}a 69,~~PL--00--681 Warszawa, ~Poland}}

\vspace{0.3cm}

{\centerline{\bf Abstract}}

\vspace{0.2cm}

Basing on the formalism of {\it generalized Dirac equation} introduced by the author already many 
years ago, we suggest a model of formal intrinsic interactions within leptons and quarks of three
generations. In this model, the leptons and quarks are treated as some intrinsic composites of 
{\it algebraic partons}, the latter being identified with Dirac bispinor indices appearing in the 
generalized Dirac equation. This intrinsic model implies a {\it universal} mass formula for leptons 
and quarks of three generations that has been recently proposed on an essentially empirical level.

\vspace{0.6cm}
 
\ni PACS numbers: 12.15.Ff , 14.60.Pq 

\vspace{0.6cm}

\ni April 2006

\vfill\eject

\baselineskip 0.76cm

\pagestyle {plain}

\setcounter{page}{1}

\vspace{0.2cm}

\ni {\bf 1. Introduction to the generalized Dirac equation}

\vspace{0.2cm}

As is well known,  the electron, the prototype of all spin-1/2 fermions, is described by the Dirac equation discovered through the famous Dirac square-root procedure

\vspace{-0.2cm}

\begin{equation}
\sqrt{p^2} \rightarrow \gamma \cdot p \,,
\end{equation}

\ni where $\gamma^\mu $ are $4\times 4$ matrices satisfying the Dirac algebra

\begin{equation}
\{\gamma^\mu, \gamma^\nu \} = 2 g^{\mu\nu}. 
\end{equation}

More than a decade ago we have observed [1] that for any positive integer $N = 1,2,3,\ldots $ there exists a more general square-root procedure

\vspace{-0.2cm}

\begin{equation}
\sqrt{p^2} \rightarrow \Gamma^{(N)}\cdot p \,,
\end{equation}

\ni where $\Gamma^{(N)\mu}$ are $4^N\times 4^N$ matrices  of the form

\vspace{-0.2cm}

\begin{equation}
\Gamma^{(N)\,\mu} \equiv \frac{1}{\sqrt{N}} \sum^N_{n=1} \gamma^{(N)\,\mu}_{n} 
\end{equation}

\ni with $\gamma^{(N)\,\mu}_{n}\; (n = 1,2,\dots,N)$ being $4^N\times 4^N$ matrices defined by the simple Clifford algebra

\begin{equation}
\left\{ \gamma^{(N)\,\mu}_{n}\,,\,\gamma^{(N)\,\nu}_m  \right\} = 2 g^{\mu \nu} \delta_{n m}\;\;\;(n,m = 1,2,...,N) \,. 
\end{equation}

\ni In fact, we check readily that such $\Gamma^{(N)\,\mu}$ matrices satisfy the Dirac algebra

\begin{equation}
\left\{ \Gamma^{(N)\,\mu}\,,\,\Gamma^{(N)\,\nu} \right\} =  2 g^{\mu \nu}\,.
\end{equation}

\ni Therefore, for any $N = 1,2,3,...$, the following {\it generalized Dirac equation} can be constructed through the procedure (3):

\begin{equation}
\left( \Gamma^{(N)} \cdot p - M^{(N)} \right) \psi^{(N)}(x) = 0 \,,
\end{equation}

\ni where the wave function $\psi^{(N)}_{\alpha_1 \alpha_2... \alpha_N}(x)$ carries $N$ Dirac bispinor indices $\alpha_n = 1,2,3,4 \;(n=1,2,...,N)$, presented in a natural way by means of the chiral representation of $N$ Clifford matrices $\gamma^{(N) \mu}_n$ (note that all matrices $\gamma^{(N) 5}_n \equiv \gamma^{(N) 0}_n\gamma^{(N) 1}_n\gamma^{(N) 2}_n\gamma^{(N) 3}_n$ and $\sigma^{(N) 3}_n \equiv \sigma^{(N) 12}_n \equiv (i/2)[\gamma^{(N) 1}_n\,,\, \gamma^{(N) 2}_n] \;(n = 1,2,...,N)$ commute).

Introducing some gauge interactions as {\it e.g.} those in the \SM, we obtain the generalized Dirac equation of the form

\begin{equation}
\left\{ \Gamma^{(N)} \cdot \left[ p - g A(x)\right] - M^{(N)} \right\} \psi^{(N)}(x) = 0 \,,
\end{equation}

\ni where $ g \Gamma^{(N)}\cdot A(x)$ symbolizes the gauge coupling (note that in the case of \SM the gauge coupling in Eq. (8) includes the matrix $\Gamma^{(N) 5} \equiv \Gamma^{(N)\,0} \Gamma^{(N) 1} \Gamma^{(N)\,2} \Gamma^{(N)\,3}$).

For $N = 1$ the generalized Dirac equation (7) is evidently the usual Dirac equation, for $N = 2$ it is known as the Dirac form [2] of K\"{a}hler equation [3], while for $N \geq 3$ (in particular for $N = 3,4,5$) it becomes a new equation. It describes a spin-halfinteger or spin-integer particle when $N$ is odd or even, respectively.

For instance, in the case of $N = 3$, the Clifford matrices $\gamma^{(N) \mu}_n$ satisfying the anticommutation relations (5) may be represented in the form:

\vspace{-0.2cm}

\begin{eqnarray}
\gamma_1^{(3) \mu} & = &  \gamma^\mu \otimes {\bf 1}\otimes {\bf 1} \;, \nonumber \\ 
\gamma_2^{(3) \mu} & = &  \gamma^5 \otimes i\gamma^\mu \gamma^5 \otimes {\bf 1} \;,\nonumber \\ \gamma_3^{(3) \mu} & = & \gamma^5 \otimes \gamma^5 \otimes \gamma^\mu  \;, 
\end{eqnarray}

\ni where $\gamma^\mu , \gamma^5 \equiv \gamma^0 \gamma^1 \gamma^2 \gamma^3$ and {\bf 1} are the usual $4\times 4$ Dirac matrices. Then, the definitions $\gamma^{(N) 5}_n \equiv 
\gamma^{(N)\,0}_n \gamma^{(N) 1}_n \gamma^{(N) 2}_n \gamma^{(N) 3}_n$  and $\sigma_n^{(N) \mu \nu} \equiv (i/2)[ \gamma^{(N) \mu}_n, \gamma^{(N) \nu}_n]$ give

\vspace{-0.2cm}

\begin{eqnarray}
\gamma^{(3) 5}_1 =  \gamma^5 \otimes {\bf 1}\otimes {\bf 1} & , & \sigma^{(3) \mu \nu}_1 = \sigma^{\mu \nu} \otimes {\bf 1}\otimes {\bf 1} \,,
\nonumber \\ 
\gamma^{(3) 5}_2  = {\bf 1} \otimes \gamma^5 \otimes {\bf 1} & , & \sigma^{(3) \mu \nu}_2 = {\bf 1}\otimes \sigma^{\mu \nu} \otimes {\bf 1} \,,
\nonumber \\ 
\gamma^{(3) 5}_3  = {\bf 1}\otimes {\bf 1} \otimes \gamma^5 & , & \sigma^{(3) \mu \nu}_3 = {\bf 1}\otimes {\bf 1}\otimes \sigma^{\mu \nu} \,,
\end{eqnarray}

\ni where $ \sigma^{\mu \nu} \equiv (i/2)[\gamma^{\mu}, \gamma^{\nu}]$ are the usual $4\times 4$ Dirac spin-tensor matrices. Analogically, we may proceed for any $N$.

The Dirac-type matrices $\Gamma^{(N)\,\mu}$ appearing in the generalized Dirac equation (7) for any $N$ can be embedded into the new Clifford algebra

\vspace{-0.2cm}

\begin{equation}
\left\{ \Gamma^{(N) \mu}_n\, ,  \,\Gamma^{(N) \nu}_m \right\} = 2g^{\mu \nu} \delta_{n m} \;\;\;(n,m = 1,2,...,N) \,,
\end{equation}

\ni isomorphic to the previous Clifford algebra (5), where the new elements $\Gamma^{(N)\,\mu}_n$ are defined by the properly normalized Jacobi linear combinations of the previous elements $\gamma^{(N)\,\mu}_n$,

\begin{eqnarray}
\Gamma^{(N) \mu}_1 & \equiv & \frac{1}{\sqrt{N}} \left(\gamma^{(N) \mu}_{1} + ... + \gamma^{(N) \mu}_{N} \right) \equiv \Gamma^{(N) \mu}\,, \nonumber \\ 
\Gamma^{(N) \mu}_{n} & \equiv & \frac{1} {\sqrt{n(n - 1)}} \left[ \gamma^{(N) \mu}_{1} + ... + \gamma^{(N) \mu}_{n\!-\!1} - (n - 1) \gamma^{(N) \mu}_n \right]\;\; (n = 2,...,N)\;.
\end{eqnarray}

\ni Then, $\Gamma^{(N) \mu}_1 \equiv \Gamma^{(N) \mu}$ and $\Gamma^{(N) \mu}_2, ..., \Gamma^{(N) \mu}_N$ are the "centre-of-mass"\,and "relative"\, Clifford matrices, respectively. Note that the generalized Dirac equation (7) involves only the "centre-of-mass"\,Clifford matrices,  being equal to the "centre-of-mass"\,Dirac-type matrices.

In place of  the chiral representation of individual Clifford matrices $\gamma^{(N) \mu}_n$, it is convenient for any $N$ to apply the chiral representation of Jacobi-type Clifford matrices $ \Gamma^{(N)\,\mu}_n$ (note that all matrices $\Gamma^{(N) 5}_n \equiv \Gamma^{(N)\,0}_n \Gamma^{(N)\,1}_n \Gamma^{(N)\,2}_n \Gamma^{(N)\,3}_n$ and $\Sigma^{(N) 3}_n \equiv 
\Sigma^{(N) 12}_n \equiv (i/2)\left[ \Gamma^{(N)\,1}_n\,,\,\Gamma^{(N)\,2}_n \right]\;(n=1,2,...,N)$ commute). Then, in the reduced representation

\begin{equation}
\Gamma^{(N) \mu} \equiv \Gamma^{(N) \mu}_1 =  \gamma^\mu \otimes \underbrace{{\bf 1}\otimes \cdots \otimes {\bf 1}}_{N-1 \;{\rm times}}  \;,
\end{equation}

\ni where $\gamma^\mu$ and {\bf 1} are the usual $4\times 4$ Dirac matrices, the generalized Dirac equation (7) takes the form

\begin{equation}
\left( \gamma \cdot p - M^{(N)}\right)_{\alpha_1\beta_1} \psi^{(N)}_{\beta_1 \alpha_2 ... \alpha_N}(x) = 0
\end{equation}

\ni with $\alpha_1$ and $\alpha_2 ,..., \alpha_N$ being the "centre-of-mass"\,and \,"relative"\, Dirac bispinor indices, respectively, where $\alpha_n = 1,2,3,4 \;(n = 1,2,...,N)$. In equation (14), $p$ and $x$ play the roles of the "centre-of-mass"\,four-momentum and "centre-of-mass"\, 
four-position, respectively. 

In the case of gauge interactions, where $p$ is replaced by $p - g A(x)$, the generalized Dirac equation becomes

\begin{equation}
\left\{ \gamma \cdot [p - g A(x)]  - M^{(N)}\right\}_{\alpha_1\beta_1} \psi^{(N)}_{\beta_1 \alpha_2 ... \alpha_N}(x) = 0 
\end{equation}

\ni with $g\gamma \cdot A(x)$ symbolizing the gauge coupling (in the case of \SM the gauge coupling in Eq. (15) includes the matrix $\Gamma^{(N) 5} \equiv  \Gamma^{(N) 5}_1$). Note that in the generalized Dirac equation (15) the \,"relative"\, bispinor indices $\alpha_2 , ..., \alpha_N$ are free from any coupling: only the "centre-of-mass"\, bispinor index $\alpha_1$ is coupled to the particle's four-momentum and to gauge fields. But, still, the \,"relative"\, bispinor indices $\alpha_2 , ..., \alpha_N$ are subject to Lorentz transformations, generated here by the total spin tensor

\begin{equation}
\sum^N_{n=1}  \sigma^{(N) \mu \nu}_n = \sum^N_{n=1}  \Sigma^{(N) \mu \nu}_n  \,,
\end{equation}

\ni where $\sigma^{(N) \mu \nu}_n \equiv (i/2)[ \gamma^{(N) \mu}_n, \gamma^{(N) \nu}_n ]$ and $\Sigma^{(N) \mu \nu}_n \equiv (i/2)[ \Gamma^{(N) \mu}_n, \Gamma^{(N) \nu}_n ]$.

We can see that the "centre-of-mass" \,bispinor index $\alpha_1$ is physically distinguished from the  "relative" \,bispinor indices $\alpha_2,...,\alpha_N$ through its coupling both to the 
four-momentum $p_\mu $ and to gauge fields $A_\mu(x)$. Thus, it is physically associated with the 
gauge-interaction charges carried by the particle as a whole (as {\it e.g.} $SU(3)\times SU(2)\times U(1)$ charges in the Standard Model). In contrast, the "relative" \,bispinor  indices $\alpha_2, ..., \alpha_N$ cannot be physically distinguished from each other as they are not experimentally observable through the gauge interactions.

Assuming that the "relative" \,bispinor  indices $\alpha_2, ..., \alpha_N$, physically undistinguishable, obey Fermi statistics along with Pauli principle (realized {\it intrinsically} for bispinor indices treated as physical objects), we impose on them full antisymmetry in the wave function $\psi^{(N)}_{\alpha_1 \alpha_2 ... \alpha_N}(x)$ [1]. Of course, the distinguished bispinor index $\alpha_1$ is apart from Fermi statistics. Since $\alpha_n$ for any $n = 1,2,...,N$ can take four values $\alpha_n = 1,2,3,4$, the number $N - 1$ of bispinor indices $\alpha_2,...,\alpha_N$, fully antisymmetric in the wave function, cannot exceed the value $N-1 = 4$. Then, only  $N = 1,2,3,4,5$ is allowed as the number of all bispinor indices in the wave function $\psi^{(N)}_{\alpha_1 \alpha_2 ... \alpha_N}(x)$. Thus, {\it only three generations} $N = 1,3,5$ of spin-1/2 particles and {\it only two generations} $N = 2,4$ of spin-0 particles can exist in the case of our generalized Dirac equation. Hence the suggestion that this equation together with the intrinsic Pauli principle are to be considered as the adequate explanation of the phenomenon of three fundamental fermion generations in the framework of \SM [1].

In this case, the wave functions of leptons and quarks of three generations $N = 1,3,5$ can be written down in terms of three wave functions $\psi^{(1,3,5)}_{\alpha_1 \alpha_2 ... \alpha_N}(x)$ in the following way [1]: 

\vspace{-0.2cm}

\begin{eqnarray} 
\psi^{(f_1)}_{\alpha_1}(x) & = & \psi^{(1)}_{\alpha_1}(x) \;, \nonumber \\
\psi^{(f_2)}_{\alpha_1}(x) & = & \frac{1}{4}\left(C^{-1} \gamma^5 \right)_ {\beta_2 \beta_3} \psi^{(3)}_{\alpha_1 \beta_2 \beta_3}(x) =\psi^{(3)}_{\alpha_1 1 2}(x) = \psi^{(3)}_{\alpha_1 3 4}(x) \,, \nonumber \\
\psi^{(f_3)}_{\alpha_1}(x) & = & \frac{1}{24}\varepsilon_{\beta_2 \beta_3 \beta_4 \beta_5} \psi^{(5)}_{\alpha_1 \beta_2 \beta_3 \beta_4 \beta_5}(x) = \psi^{(5)}_{\alpha_1 1 2 3 4}(x) \;,
\end{eqnarray}  

\ni where $ \psi^{(N)}_{\alpha_1 \alpha_2 ... \alpha_N}(x) \;(n=1,3,5)$ carry also the \SM (composite) label suppressed in our notation, while $C$ is the usual $4\times 4$ 
charge-conjugation matrix.  Here, $f_1 = \nu_e, e^-, u, d \;,\; f_2 = \nu_\mu, \mu^-, c, s $ and $f_3 = \nu_\tau, \tau^-, t , b $. Of course, four kinds of fundamental fermions: neutrinos, charged leptons, up quarks and down quarks differ by the suppressed \SM (composite) label. 

It can be seen that (due to the full antisymmetry of $\alpha_2, ..., \alpha_N $ indices) the wave functions or fields (17), corresponding to $N =1,3,5$, appear (up to the sign $\pm$) with the multiplicities 1, 4, 24,  respectively. So, there are defined here the {\it generation-weighting factors} [1,4]

\begin{equation} 
\rho_1 = \frac{1}{29} \;,\; \rho_2 = \frac{4}{29} \;,\; \rho_3 = \frac{24}{29}
\end{equation}

\ni ($\sum_i\rho_i = 1$), such that the bilinear relation

\vspace{-0.2cm}

\begin{equation}
\psi^{(N_i)\dagger}(x) \psi^{(N_i)}(x) = 29 \rho_i \psi^{(f_i)\dagger}(x) \psi^{(f_i)}(x)  
\end{equation}

\ni holds for three numbers of Dirac bispinor indices

\begin{equation}
N_1 = 1 \;,\; N_2 = 3 \;,\; N_3 = 5
\end{equation}

\ni numerating three generations $i = 1,2,3$.


\vspace{0.2cm}

\ni {\bf 2. Construction of the universal mass formula}

\vspace{0.2cm}

In order to construct in our formalism a mass formula for leptons and quarks of three generations, we get to our disposal the numbers $N_i = 1,3,5$ and $\rho_i = 1/29,4/29, 24/29$ corresponding to three generations $i =1,2,3$ and, of course, some constants taking values corresponding to four kinds of fundamental fermions: neutrinos, charged leptons, up quarks, down quarks, labelled by $f = \nu, l, u, d$ (with $u$ and $d$ denoting here all up and down quarks). On a phenomenological level, these constants may be treated as free parameters to be determined from experimental data.

We will look for the mass formula in the form

\vspace{-0.2cm}

\begin{equation}
m_{f_i} = \rho_i  h_{f_i} \;\;\;(i = 1,2,3)\,,
\end{equation}

\ni where $h_{f_i}$ for any $f \!=\! \nu, l, u, d$ is a linear combination of three terms $u_i, v_i, w_i$ to be discovered:

\vspace{-0.2cm}

\begin{equation}
h_{f_i} = a^{(f)} u_i + b^{(f)} v_i + c^{(f)} w_i  \;\;\; (i =1,2,3)\;.
\end{equation}

\ni Here, $a^{(f)}, b^{(f)}, c^{(f)}$ are constants dependent on the \SM (composite) label determining four kinds $f$ of fundamental fermions. Generically, Eqs. (21) with (22) give for any kind $f$ a linear transformation of three masses $m_{f_1}, m_{f_2}, m_{f_3}$ into three free parameters $a^{(f)}, b^{(f)}, c^{(f)}$. The terms $u_i, v_i, w_i \; (i=1,2,3)$ describe nine coefficients which are {\it universal} for all four kinds $f = \nu, l, u, d$ of fundamental fermions. If the parameters $a^{(f)}, b^{(f)}, c^{(f)}$ are dependent, then Eqs. (21) with (22) give some {\it predictions} for the masses $m_{f_1}, m_{f_2}, m_{f_3}$ (of course, when the terms $u_i, v_i, w_i \; (i=1,2,3)$ are known).

Since in our formalism the Dirac bispinor indices $\alpha_n = 1,2,3,4 \;\;(n = 1,2,..., N_i$ and $N_i = 1,3,5)$ are considered as physical objects (call them {\it algebraic partons} of leptons and quarks), we will try to connect the terms $u_i, v_i, w_i$ in Eq. (22) (multiplied by $a^{(f)}, b^{(f)}, c^{(f)}$, respectively) with some formal intrinsic interactions that, intuitively, may act on the algebraic partons within leptons and quarks{\footnote{It may be that the above intrinsic model of leptons and quarks (with Dirac bispinor indices playing the role of algebraic partons) shows up only the summit of a hidden iceberg of a spatial model, where the fundamental fermions of three generations $i=1,2,3$ consist of $N_i=1,3,5$ spin-1/2 spatial partons getting both Dirac bispinor indices $\alpha_1,\alpha_2,...,\alpha_N$ and positions $\vec{x}_1,\vec{x}_2,...,\vec{x}_{N_i}$ (or momenta $\vec{p}_1,\vec{p}_2,...,\vec{p}_{N_i}$), bound mainly in orbital $S$ states. One of these partons ought to carry also the set of \SM charges, while the rest of them ought to be  \SM sterile (forming a new atom-like system). Then, of course, some binding forces of a new nature are needed. At present, the lack of any adequate proposal of such binding forces is in favor of an intrinsic model based on our generalized Dirac equation.}}.

The first candidate for such interactions may be a two-body intrinsic interaction between all $N_i$ algebraic partons $n = 1,2,...,N_i$ treated on an equal footing:

\vspace{-0.2cm}

\begin{equation}
u_i = \sum_{n,m =1}^{N_i} 1 = N_i^2 = N_i + N_i(N_i-1) \;\;\; (i = 1,2,3)\,.
\end{equation}

\ni The part $\sum_{n = m} 1 = N_i$ of the term (23) (multiplied by $a^{(f)} > 0$) describes $N_i$ intrinsic self-interactions ($N_i$ intrinsic masses), each equal to $a^{(f)}$. The part $\sum_{n \neq m} 1 = N_i(N_i-1)$ of the term (23) (multiplied by $a^{(f)} > 0$) gives $N_i(N_i - 1)$ intrinsic interactions of different algebraic partons, each equal to $a^{(f)}$. Vanishing for the generation $i=1$, it allows together with the part $\sum_{n = m} 1 = N_i$ to interpret the generations  $i=2$ and 3 as two intrinsic excitations of the ground generation $i=1$, rising faster than $N_i$ (remember also the additional rising factor $\rho_i$ in Eq. (21)). Notice that the sum $\sum_{n \neq m} 1 = N_i$ contains terms both with $n<m$ and $n>m$: $\sum_{n < m} 1 + \sum_{n > m} 1 = N_i(N_i -1)$, what is in spirit of our equal treating of all algebraic partons, assumed in Eq. (23) (and then in Eq. (25)).

The second candidate may correct the intrinsic self-interaction (intrinsic mass) of the algebraic parton $n=1$ (distinguished from all other $N_i -1$ partons that are undistinguishable from each other and obey the intrinsic Pauli principle):

\begin{equation}
v_i = \left(P_{n =1}^{(N_i)}\right)^2 = \left[\frac{N_i!}{(N_i-1)!}\right]^{-2} = \frac{1}{N_i^2} \;\;\; (i = 1,2,3)\,.
\end{equation}

\ni Here, $P^{(N_i)}_{n =1} = [N_i!/(N_i-1)!]^{-1} = 1/N_i$ describes the probability of finding the distinguished algebraic parton among all $N_i$ partons of which $N_i-1$ are undistinguishable. Note that $ P^{(N_i)}_{n \neq 1} = 1 - P^{(N_i)}_{n =1} = (N_i -1)/N_i $ is the probability of finding {\it any} of undistinguishable algebraic partons {\it i.e.}, $n = 2$ or 3 ... or $N_i$. The resulting intrinsic mass of the algebraic parton $n=1$ is $a^{(f)} + b^{(f)}/N_i^2$.

Finally, the third candidate may be a collective correction to the intrinsic self-interactions (intrinsic masses) of all $N_i$ algebraic partons $n = 1,2,\dots,N_i$ treated on an equal footing:

\begin{equation}
w_i =  -\left(P^{(N_i)}_{n=1} +P^{(N_i)}_{n \neq 1}\right) = -1 \;\;\; (i = 1,2,3)\,.
\end{equation}

\ni Due to the minus sign, this term (when multiplied by $c^{(f)} > 0$) gets the form of mass deficit that binds $N_i$ intrinsic masses appearing in the term (23) (multiplied by $a^{(f)} > 0$), as if they were parts of a spatial bound system. The resulting effective intrinsic mass of all algebraic partons is $a^{(f)} N_i + b^{(f)}/N_i^2 - c^{(f)}$. To this effective mass, the intrinsic interaction $a^{(f)}N_i(N_i-1)$ of different partons ought to be added within a lepton or quark.

Then, with the notation 

\begin{equation}
a^{(f)} = \mu^{(f)} >0 \,,\, b^{(f)} = \mu^{(f)}(\varepsilon^{(f)} - 1) \,,\, c^{(f)}= \mu^{(f)}\xi^{(f)} >0 \,,
\end{equation}

\ni convenient in our further discussion, we obtain from Eq. (22) the linear combinations

\begin{equation}
h_{f_i}  =  \mu^{(f)} \left(N^2_i + \frac{\varepsilon^{(f)} -1}{N^2_i} - \xi^{(f)}\right) \:\;\;(i = 1,2,3)\,.
\end{equation}

\ni Thus, in the considered model of $u_i, v_i, w_i$, the mass formula (21) for any $ f = \nu, l,u,d$ takes the following specific shape [5]:

\begin{equation}
m_{f_i}  =  \mu^{(f)}\, \rho_i \left(N^2_i + \frac{\varepsilon^{(f)} -1}{N^2_i} - \xi^{(f)}\right) \;\;(i = 1,2,3)\;.
\end{equation}

When written down explicitly, this gives

\begin{eqnarray}
m_{f_1} & = & \frac{\mu^{(f)}}{29} (\varepsilon^{(f)} - \xi^{(f)}) \,, \nonumber \\
m_{f_2} & = & \frac{\mu^{(f)}}{29} \frac{4}{9} (80 +\varepsilon^{(f)} - 9\,\xi^{(f)}) \,, \nonumber \\
m_{f_3} & = & \frac{\mu^{(f)}}{29} \frac{24}{25} (624 + \varepsilon^{(f)} - 25\,\xi^{(f)}) \,.
\end{eqnarray}

It can be seen that the mass formula (28) has a {\it universal} shape for all leptons and quarks: it is of the form $ m_{f_i} = F_i(\mu^{(f)}, \varepsilon^{(f)}, \xi^{(f)})$, where $F_i$ (depending on generations $i=1,2,3$) is independent of kinds $f = \nu, l, u, d$ of fundamental fermions. Of course, through the mass formula (28), twelve masses $m_{f_i}$ (if all were known) could determine twelve parameters $\mu^{(f)}, \varepsilon^{(f)}, \xi^{(f)}$ or {\it vice versa}. If parameters are dependent, some masses can be {\it predicted}.

Strictly speaking, in the case of mass neutrinos $\nu_i$, the mass formula (28) ought to work for Dirac neutrino masses $m^{(D)}_{\nu_i}$. If the effective active-neutrino masses $m_{\nu_i}$ can be evaluated from the simple seesaw formula 

\begin{equation} 
m_{\nu_i} = - \frac{ m_{\nu_i}^{(D)\,2}}{M_{\nu_i}} \,,
\end{equation}

\ni and if the heavy Majorana sterile-neutrino masses $M_{\nu_i}$ satisfy the proportionality relation

\begin{equation} 
M_{\nu_i} = \zeta m_{\nu_i}^{(D)} 
\end{equation}

\ni containing a very large coefficient $\zeta >0$, then $m_{\nu_i} = -m^{(D)}_{\nu_i}/\xi $ and the active-neutrino mass formula gets the following specific shape [5]: 

\begin{equation}
m_{\nu_i}  = \frac{\mu^{(\nu)} \,\xi^{(\nu)}}{\zeta} \rho_i \left[1- \frac{1}{\xi^{(\nu)}} \left(N^2_i + \frac{\varepsilon^{(\nu)} -1}{N^2_i}\right)\right] \;\;(i = 1,2,3)\;.
\end{equation}

\ni Here, $\mu^{(\nu)}\,\!' \equiv \mu^{(\nu)} \,\xi^{(\nu)}/\zeta \,,\, \varepsilon^{(\nu)}\,\!' \equiv \varepsilon^{(\nu)}/\xi^{(\nu}) $ and $1/\xi^{(\nu)}$ are three new parameters. 

The mass formula (28) for leptons and quarks (and (32) in the case of active neutrinos) has been introduced and discussed in Ref. [5] on an essentially empirical level. In the present note, this formula is intuitively justified on the ground of our formalism of generalized Dirac equation.

\vspace{0.2cm}

\ni {\bf 3. The best test}

\vspace{0.2cm}

The best experimental test of the mass formula (28) works for charged leptons $l_i = e^-, \mu^-, \tau^-$, where [6]

\begin{equation}
m_{e} = 0.5109989\;{\rm MeV} \;,\; m_{\mu} = 105.65837\;{\rm MeV} \;,\; m_{\tau} = 1776.99^{+0.29}_{-0.26}\;{\rm MeV} \;.
\end{equation}

\ni In this case, the values of parameters $\mu^{(l)}$ and $\varepsilon^{(l)}$ in Eq. (28) are [5]

\begin{equation}
\mu^{(l)}  = 86.0076\;{\rm MeV} \;,\;\varepsilon^{(l)} = 0.174069 \;,
\end{equation}

\ni and then

\begin{equation}
\varepsilon^{(l)} - \xi^{(l)} = \frac{29 m_e}{\mu^{(l)}} = 0.172298 
\end{equation}

\ni and 

\vspace{-0.2cm}

\begin{equation}
\xi^{(l)} \equiv \varepsilon^{(l)} - (\varepsilon^{(l)} - \xi^{(l)}) = 1.771\times 10^{-3} =1.8\times 10^{-3} \,.
\end{equation}

\ni Here, for $m_\tau$ its central experimental value given in Eq. (33) is used. Thus, the charged-lepton parameter $\xi^{(l)}$ is small in comparison with the terms $N^2_i + (\varepsilon^{(l)}-1)/N^2_i $ in Eq. (28).

If we put exactly $\xi^{(l)} = 0$, we would predict from Eqs. (29) [1,5]

\begin{equation} 
m_\tau = \frac{6}{125} (351m_\mu -  136 m_e) = 1776.80\;{\rm MeV} \,,
\end{equation}

\ni and would evaluate that

\begin{equation}
\mu^{(l)} = \frac{29 (9m_\mu - 4 m_e)}{320} = 85.9924\;{\rm MeV}
\end{equation}

\ni and

\begin{equation}
\varepsilon^{(l)} = \frac{320 m_e}{9 m_\mu - 4 m_e} = 0.172329\,.
\end{equation}

\ni Here, only the input of experimental $m_e$ and $m_\mu$ is applied.

We can see that the prediction (37) for $m_\tau $, valid in the limit of $\xi^{(l)}\rightarrow 0$, is in a very good agreement with experimental $m_\tau $ given in Eq. (33). But, the small parameter $\xi^{(l)}$ is probably not exactly zero, since the central experimental value of $m_\tau$ is 1776.99 MeV corresponding to $\xi^{(l)}= 1.8\times 10^{-3}$. Nevertheless, this impressive agreement (even not fully exact) is a strong support for our specific mass formula (28), at least in the case of charged leptons.

\vspace{0.2cm}

\ni {\bf 4. Other consequences}

\vspace{0.2cm}

It turns out that, in contrast to the small charged-lepton parameter $\xi^{(l)}$, the neutrino parameter $\xi^{(\nu)}$ is large. In fact, making use of the experimental estimates [7]

\begin{equation}
|m^2_{\nu_2} - m^2_{\nu_1}| \sim 8.0\times 10^{-5}\; {\rm eV}^2 \;,\; |m^2_{\nu_3} - m^2_{\nu_2}| \sim 2.4\times 10^{-3}\; {\rm eV}^2 \;,
\end{equation}

\ni giving the ratio $|m^2_{\nu_3} - m^2_{\nu_2}| /|m^2_{\nu_2} - m^2_{\nu_1}| \sim 30$, and considering the normal hierarchy $ m^2_{\nu_1}\ll m^2_{\nu_2} \ll m^2_{\nu_3}$ with the lowest mass lying in the range

\begin{equation}
m_{\nu_1} \sim(0 \;{\rm to} \;10^{-3})\;{\rm eV} \;,
\end{equation}

\ni we obtain [5]

\begin{equation}
m_{\nu_2} \sim(8.9 \;{\rm to} \;9.0)\times 10^{-3}\;{\rm eV} \;,\;m_{\nu_3} \sim 5.0\times 10^{-2}\;{\rm eV} \;.
\end{equation}

\ni Then, the values of parameters in Eq. (32) are [5]:

\begin{equation}
\frac{\mu^{(\nu)} \xi^{(\nu)}}{\zeta} \sim (7.9\;{\rm to} \;7.5)\times 10^{-2}\;{\rm eV} \;,\; \frac{\varepsilon^{(\nu)}}{ \xi^{(\nu)}} \sim 1\;{\rm to} \;0.61
\end{equation}

\ni and

\begin{equation}
\frac{1}{\xi^{(\nu)}} \sim (8.1\;{\rm to} \;6.9)\times 10^{-3} \;.
\end{equation}

\ni So, in this case, the neutrino parameter $1/\xi^{(\nu)}$ is small in Eq. (32), and $\varepsilon^{(\nu)}/\xi^{(\nu)} $ is there not very different from 1.

The corresponding seesaw parameter $\zeta $  becomes

\begin{equation}
\zeta \equiv \frac{\mu^{(l)}}{\mu^{(\nu)}/\zeta}\,\frac{\mu^{(\nu)}}{\mu^{(l)}} \sim (1.3 \;{\rm to}\; 1.7)\times 10^{11}\, \frac{\mu^{(\nu)}}{\mu^{(l)}} \,,
\end{equation}

\ni where $\mu^{(l)} = 86.0076$ MeV (Eq. (34)) and $\mu^{(\nu)}/\zeta \sim (6.4\;{\rm to}\; 5.2)\times 10^{-4}$ eV (Eqs. (43) and (44)). If {\it e.g.} $\mu^{(\nu)} \sim \mu^{(l)}$, then $\zeta \sim O(10^{11})$.

If we put in Eq, (32) exactly $\varepsilon^{(\nu)}/\xi^{(\nu)} = 1$, we would {\it predict} $m_{\nu_1} = 0$, implying from the estimates (40) that $m_{\nu_2} \sim 8.9\times 10^{-3}$ eV and $m_{\nu_3} \sim 5.0\times 10^{-2}$ eV, and then would evaluate that  $\mu^{(\nu)} \xi^{(\nu)}/\zeta \sim 7.9\times 10^{-2}$ eV and $ 1/\xi^{(\nu)} \sim 8.1\times 10^{-3}$.

If we put there exactly $1/\xi^{(\nu)} = 0$, we would obtain quite different figures. In fact, using in this case the experimental estimates (40), we would {\it predict} from Eq. (32) [5]

\begin{equation}
m_{\nu_1} \sim 1.5\times 10^{-2}\;{\rm eV}\;,\;m_{\nu_2} \sim 1.2\times 10^{-2}\;{\rm eV}\;,\; m_{\nu_3} \sim 5.1\times 10^{-2}\;{\rm eV}\;,
\end{equation}

\ni where 

\begin{equation}
m_{\nu_3} = \frac{6}{25}\left(27m_{\nu_2} - 8 m_{\nu_1}\right) \,,
\end{equation}

\ni and would evaluate that

\begin{equation}
\frac{\mu^{(\nu)} \xi^{(\nu)}}{\zeta} = \frac{29}{32}(9 m_{\nu_2} - 4 m_{\nu_1}) \sim 4.5\times 10^{-2}\;{\rm eV}  \;,\; \frac{\varepsilon^{(\nu)}}{\xi^{(\nu)}} = 1 - \frac{32 m_{\nu_1}}{9 m_{\nu_2} - 4 m_{\nu_1}} \sim -8.8 
\end{equation}

\ni (to be more precise, we would {\it predict} one of three masses (46), say $m_{\nu_1}$). 

We can see from the estimates (46) that in the limit of $1/\xi^{(\nu)} \rightarrow 0$ the mass ordering of neutrino states 1 and 2 is {\it inverted}, while the position of neutrino state 3 is {\it normal}. Note also the minus sign of $\varepsilon^{(\nu)}$, in contrast to its plus sign in the case of $m_{\nu_1} \stackrel {<}{\sim} \,10^{-3}$~eV, where the hierarchy is normal and the small parameter $1/\xi^{(\nu)}$ is not exactly zero.

For up and down quarks, making use of medium experimental mass estimates [6]

\begin{equation} 
m_u \sim 2.8 \;{\rm MeV}\;,\;  m_c \sim 1.3 \;{\rm GeV}\;,\;  m_t \sim 174 \;{\rm GeV} 
\end{equation}

\ni and

\begin{equation} 
m_d \sim 6 \;{\rm MeV}\;,\;  m_s \sim 110 \;{\rm MeV}\;,\;  m_b \sim 4.3\;{\rm GeV} \,,
\end{equation}

\ni we evaluate the parameters in Eq. (28) [5]:

\begin{equation} 
\mu^{(u)} \sim 13 \;{\rm GeV}\;,\; \varepsilon^{(u)} \sim 9.2 \;,\; \xi^{(u)} \sim 9.2
\end{equation}

\ni and

\begin{equation} 
\mu^{(d)} \sim 280 \;{\rm MeV}\;,\; \varepsilon^{(d)} \sim 7.5 \;,\; \xi^{(d)} \sim 6.9 
\end{equation}

\ni and, more precisely,

\begin{equation} 
\varepsilon^{(u)} - \xi^{(u)} = \frac{29 m_u}{\mu^{(u)}} \sim 6.2\times 10^{-3} \;\;,\;\;\varepsilon^{(d)} - \xi^{(d)} = \frac{29 m_d}{\mu^{(d)}} \;\sim\; 6.1\times 10^{-1}\,.
\end{equation}

\ni Notice that the up-quark value $\varepsilon^{(u)} - \xi^{(u)}$ is small and the down-quark value  $\varepsilon^{(d)} - \xi^{(d)}$ is not large.

If for up quarks we put exactly $\varepsilon^{(u)} - \xi^{(u)} = 0$, we would {\it predict}

\begin{equation} 
m_u = \frac{\mu^{(u)}}{29} (\varepsilon^{(u)} - \xi^{(u)}) = 0\, , 
\end{equation}

\ni and then would evaluate

\begin{eqnarray}
\mu^{(u)}& = & \frac{29\cdot 25}{1536\cdot 6}(m_t - \frac{6}{25} 27 m_c) \sim 13\;{\rm GeV} \;,\\
\varepsilon^{(u)} & = & \xi^{(u)}= 10\,\frac{m_t - \frac{6}{125} 351 m_c}{m_t - \frac{6}{25} 27 m_c} \sim 9.2
\end{eqnarray}

\vspace{0.1cm}

\ni after the input of experimental $m_c $ and $m_t $ given in Eq. (49). So, the parameters $\mu^{(u)}\,,\, \varepsilon^{(u)}$ and $ \xi^{(u)}$ do not change (on the level of two decimals) in the limit of $\varepsilon^{(u)} - \xi^{(u)} \rightarrow 0$. In this calculation, we applied Eqs. (5) and (8) -- (10) in Ref. [5] (the same equations were used to calculate the values (51), (52) and (53)).

\vspace{0.2cm}

\ni {\bf 5. Conclusions}

\vspace{0.2cm}

On grounds of the {\it generalized Dirac equation}, found several years ago in the framework of the {\it generalized Dirac's square root} [1], we have constructed a model of formal intrinsic interactions within leptons and quarks of three generations.

In this model, leptons and quarks of three generations are treated as some intrinsic composites of {\it algebraic partons} which are identified with $N = 1,3,5$ Dirac bispinor indices provided by the generalized Dirac's square root procedure. Of these $N$ indices, $N-1$ are assumed to obey the {\it intrinsic Pauli principle}: denoted by $\alpha_2,...,\alpha_N$, they are antisymmetrized in the wave functions $\psi^{(N)}_{\alpha_1 \alpha_2 ... \alpha_N}(x)$ of leptons and quarks (here, the \SM (composite) label, determining four kinds $f = \nu, l, u, d$ of leptons and quarks, is suppressed). 

Such intrinsic model leads to a {\it universal} mass formula  for leptons and quarks of three generations, recently proposed on an essentially empirical level [5]. At present, the best experimental test of this formula works successfully in the case of charged leptons $f = l$.

We hope that our model of formal intrinsic interactions within leptons and quarks might be a prototype of a future theory of pointlike but composite particles, which -- we believe -- is needed to describe the spectrum of fundamental fermions appearing in the Standard Model. However, we should still remember the option of "the summit of a hidden iceberg"\, mentioned in the footnote \dag.

Eventually, we should like to remind that in our formal model of intrinsic interactions the dependence of parameters $\mu^{(f)}, \varepsilon^{(f)}, \xi^{(f)}$ on the index $f = \nu, l, u,d$ numerating four kinds of fundamental fermions remains fully phenomenological, though it is expected to be physically determined by the \SM charges (leading to the \SM (composite) label suppressed in our notation). From the theoretical point of view, such phenomenology is perhaps the main unsatisfactory feature of our model. But, the physical explanation of this phenomenological dependence may be the most fascinating goal for a future more advanced  theory.

\vfill\eject

~~~~
\vspace{0.5cm}

{\centerline{\bf References}}

\vspace{0.5cm}

{\everypar={\hangindent=0.65truecm}
\parindent=0pt\frenchspacing

{\everypar={\hangindent=0.65truecm}
\parindent=0pt\frenchspacing

[1]~W. Kr\'{o}likowski, {\it Acta Phys. Pol.} {\bf B 21}, 871 (1990); {\it Phys. Rev.} {\bf D 45}, 3222 (1992); {\it Acta Phys. Pol.} {\bf B 33}, 2559 (2002) [{\tt hep--ph/0203107}]; {\tt hep--ph/0504256}. 

\vspace{0.2cm}

[2]~T. Banks, Y. Dothan and D.~~Horn, {\it Phys. Lett.} {\bf B 117}, 413 (1982).

\vspace{0.2cm}

[3]~E. K\"{a}hler, {\it Rendiconti di Matematica} {\bf 21}, 425 (1962); {\it cf.} also D.~Ivanenko and L.~Landau, {\it Z. Phys.} {\bf 48}, 341 (1928).

\vspace{0.2cm}

[4]~{\it Cf.} also W. Kr\'{o}likowski, {\it Acta Phys. Pol.} {\bf B 36}, 2051 (2005) 
[{\tt hep--ph/0503074}]; {\tt hep-ph/0510355}. 

\vspace{0.2cm}

[5]~W. Kr\'{o}likowski, {\tt hep--ph/0602018}. 

\vspace{0.2cm}
 
[6]~Particle Data Group, {\it Review of Particle Physics, Phys.~Lett.} {\bf B 592} (2004).

\vspace{0.2cm}

[7]~{\it Cf. e.g.} G.L. Fogli, E. Lisi, A. Marrone and A. Palazzo, {\tt hep--ph/0506083}.

\vspace{0.2cm}

\vfill\eject

\end{document}